\documentstyle[12pt]{article}
\font\tenrm=cmr10

\begin{document}
\renewenvironment{thebibliography}[1]
  { \begin{list}{\arabic{enumi}.}
    {\usecounter{enumi} \setlength{\parsep}{0pt}
     \setlength{\itemsep}{3pt} \settowidth{\labelwidth}{#1.}
     \sloppy
    }}{\end{list}}

\parindent=1.5pc

\begin{flushright} UCRHEP-T112\\June 1993\
\end{flushright}
\vglue 1.0cm
\begin{center}{{\bf NEW SUPERSYMMETRIC TWO-HIGGS-DOUBLET\\
               \vglue 3pt
               STRUCTURE AT THE ELECTROWEAK ENERGY SCALE\\}
\vglue 1.0cm
{ERNEST MA}\\
\baselineskip=14pt
{\it Department of Physics, University of California,}\\
\baselineskip=14pt
{\it Riverside, California 92521, USA}\\
\vglue 0.8cm
{ABSTRACT}}
\end{center}
\vglue 0.3cm
{\rightskip=3pc
 \leftskip=3pc
 \tenrm\baselineskip=12pt
 \noindent
Contrary to common belief, the requirement that supersymmetry exists and
that there are two Higgs doublets and no singlet at the electroweak
energy scale does not necessarily result in the minimal supersymmetric
standard model (MSSM).  An interesting alternative is presented.
\vglue 0.8cm}
{\bf\noindent 1. Introduction}
\vglue 0.4cm
\baselineskip=14pt
It is generally believed that given the gauge group SU(2) $\times$ U(1)
and the requirement of supersymmetry, the quartic scalar couplings of the
Higgs potential (consisting of two doublets and no singlet) are completely
determined in terms of the two gauge couplings.  This is actually not the
case because the SU(2) $\times$ U(1) gauge symmetry may be a remnant\cite{1}
of a larger symmetry which is broken at a higher mass scale {\underline
{together with the supersymmetry}}.  The structure of the Higgs potential
is then determined by the scalar particle content needed to precipitate
the proper spontaneous symmetry breaking and to render massive the assumed
fermionic content of the larger theory.  Furthermore, the quartic scalar
couplings are related to the gauge couplings of the larger theory
{\underline {as well as other couplings}} appearing in its superpotential.
At the electroweak energy scale, the reduced Higgs potential may contain
only two scalar doublets, but their quartic couplings may not be those of
the minimal supersymmetric standard model (MSSM).  The work that I will
describe in this contribution to Kamesh Wali's Festschrift is an explicit
first example that the MSSM structure is not unique.  It is based on my very
recent work with Daniel Ng of TRIUMF.\cite{2}
\vglue 0.6cm
{\bf\noindent 2. The Two-Doublet Higgs Potential}
\vglue 0.4cm
Consider two Higgs doublets $\Phi_{1,2} = (\phi_{1,2}^+,\phi_{1,2}^0)$ and the
Higgs potential
\begin{eqnarray}
V &=& \mu_1^2 \Phi_1^\dagger \Phi_1 + \mu_2^2 \Phi_2^\dagger \Phi_2 +
\mu_{12}^2 (\Phi_1^\dagger \Phi_2 + \Phi_2^\dagger \Phi_1) \nonumber \\
&+& {1 \over 2} \lambda_1 (\Phi_1^\dagger \Phi_1)^2 + {1 \over 2} \lambda_2
(\Phi_2^\dagger \Phi_2)^2 + \lambda_3 (\Phi_1^\dagger \Phi_1) (\Phi_2^\dagger
\Phi_2) \nonumber \\ &+& \lambda_4 (\Phi_1^\dagger \Phi_2) (\Phi_2^\dagger
\Phi_1) + {1 \over 2} \lambda_5 (\Phi_1^\dagger \Phi_2)^2 + {1 \over 2}
\lambda_5^* (\Phi_2^\dagger \Phi_1)^2.
\end{eqnarray}
In the MSSM, there are the well-known constraints
\begin{equation}
\lambda_1 = \lambda_2 = {1 \over 4} (g_1^2 + g_2^2), ~~ \lambda_3 =
- {1 \over 4} g_1^2 + {1 \over 4} g_2^2, ~~ \lambda_4 = - {1 \over 2} g_2^2,
{}~~ \lambda_5 = 0,
\end{equation}
where $g_1$ and $g_2$ are the U(1) and SU(2) gauge couplings of the standard
model respectively.  Note that only the gauge couplings contribute to the
$\lambda$'s.  This is because that with only two SU(2) $\times$ U(1) Higgs
superfields, there is no cubic invariant in the superpotential and thus no
additional coupling.
\vglue 0.6cm
{\bf \noindent 3. The E$_6$-Inspired Left-Right Model}
\vglue 0.4cm
Consider now the gauge group $\rm SU(2)_L \times SU(2)_R \times U(1)$ but with
an unconventional assignment of fermions.\cite{3}  An exotic quark $h$ of
electric charge $-1/3$ is added so that $(u,d)_L$ transforms as (2,1,1/6),
$(u,h)_R$ as (1,2,1/6), whereas both $d_R$ and $h_L$ are singlets
(1,1,$-1/3$).  There are two scalar doublets $\Phi_1$ and $\chi$, as well
as a bidoublet
\begin{equation}
\eta = \left( \begin{array} {c@{\quad}c} \overline {\phi_2^0} & \eta^+ \\
- \phi_2^- & \eta^0 \end{array} \right),
\end{equation}
transforming as (2,1,1/2), (1,2,1/2), and (2,2,0) respectively.  Note that
$\Phi_1^\dagger \tilde \eta \chi$ is then an allowed term in the
superpotential, where $\tilde \eta \equiv \sigma_2 \eta^* \sigma_2$, so
that its coupling $f$ also contributes to the quartic scalar couplings of
this model's Higgs potential.

Let $G_1$ be the U(1) gauge coupling and $G_2$ the coupling of both SU(2)'s.
Then
\begin{eqnarray}
V &=& V_{soft} + {1 \over 8} (G_1^2 + G_2^2) [(\Phi_1^\dagger \Phi_1)^2 +
(\chi^\dagger \chi)^2] \nonumber \\ &+& {1 \over 4} G_2^2 [(Tr \eta^\dagger
\eta)^2 - (Tr \eta^\dagger \tilde \eta)(Tr \tilde \eta^\dagger \eta)] +
(f^2 - {1 \over 4} G_2^2) (\Phi_1^\dagger \Phi_1 + \chi^\dagger \chi)
Tr \eta^\dagger \eta \nonumber \\ &-& (f^2 - {1 \over 2} G_2^2)
(\Phi_1^\dagger \eta \eta^\dagger \Phi_1 + \chi^\dagger \eta^\dagger \eta
\chi) + (f^2 - {1 \over 4} G_1^2) (\Phi_1^\dagger \Phi_1)(\chi^\dagger \chi),
\end{eqnarray}
where $V_{soft}$ contains terms of dimensions 2 and 3, and breaks the
supersymmetry.  Let $\chi^0$ acquire a vacuum expectation value $u \neq 0$.
Then $\rm SU(2)_L \times SU(2)_R \times U(1)$ breaks down to the standard
$\rm SU(2)_L \times U(1)_Y$ with $m^2(\sqrt 2 Re \chi^0) = (G_1^2 + G_2^2)
u^2/2$ and $m^2(\eta^+,\eta^0) = G_2^2 u^2/2$.  These heavy particles can
be integrated out at the electroweak energy scale where only $\Phi_{1,2}$
are left.
\vglue 0.6cm
{\bf \noindent 4. Reduced Higgs Potential of the Left-Right Model}
\vglue 0.4cm
The quartic scalar couplings of the reduced Higgs potential at the electroweak
energy scale are now given by
\begin{eqnarray}
\lambda_1 &=& {1 \over 4} (G_1^2 + G_2^2) - {(4f^2-G_1^2)^2 \over {4(G_1^2+
G_2^2)}}, \\ \lambda_2 &=& {1 \over 2} G_2^2 - {(4f^2 - G_2^2)^2 \over
{4(G_1^2 + G_2^2)}}, \\ \lambda_3 &=& {1 \over 4} G_2^2 - {{(4f^2-G_1^2)
(4f^2-G_2^2)} \over {4(G_1^2+G_2^2)}}, \\ \lambda_4 &=& f^2 - {1 \over 2}
G_2^2, ~~ \lambda_5 = 0,
\end{eqnarray}
where the second terms on the right-hand sides of the equations for
$\lambda_{1,2,3}$ come from the cubic interactions of $\sqrt 2 Re \chi^0$.
In the limit $f=0$ and using the tree-level boundary conditions $G_2 = g_2$
and $G_1^{-2} + G_2^{-2} = g_1^{-2}$, it can easily be shown from the above
that the MSSM is recovered.  However, $f$ is in general nonzero, although
it does have an upper bound because $V$ must be bounded from below.  Hence
\begin{equation}
0 \leq f^2 \leq {1 \over 4} (g_1^2 + g_2^2) \left( 1 - {g_1^2 \over g_2^2}
\right)^{-1},
\end{equation}
where the maximum value is obtained if $V_{soft}$ is also left-right
symmetric.
\vglue 0.6cm
{\bf \noindent 5. Phenomenological Consequences}
\vglue 0.4cm
For illustration, let $f = f_{max}$ and $x \equiv \sin^2 \theta_W$, then
\begin{equation}
\lambda_1 = 0, ~~ \lambda_2 = {e^2 \over {2x}} \left[ 1 - {{2x^2} \over
{(1-x)(1-2x)}} \right] + {{g_2^2 \epsilon} \over {4M_W^2 \sin^4 \beta}},
\end{equation}
\begin{equation}
\lambda_3 = {e^2 \over {4x}} \left[ 1 - {{2x} \over {1-2x}} \right] =
- \lambda_4, ~~ \lambda_5 = 0,
\end{equation}
where
\begin{equation}
\epsilon = {{3g_2^2m_t^4} \over {8\pi^2M_W^2}} \ln \left( 1 + {\tilde m^2
\over m_t^2} \right)
\end{equation}
is an
extra term coming from radiative corrections and $\tan \beta \equiv \langle
\phi_2^0 \rangle / \langle \phi_1^0 \rangle$.  Comparing against the MSSM,
the lighter of the two neutral scalar bosons is now constrained by
\begin{equation}
m_h^2 < m_A^2 \sin^2 \beta, ~~ m_h^2 < 2 M_W^2 \left[ 1-{{2x^2} \over
{(1-x)(1-2x)}} \right] \sin^4 \beta + \epsilon,
\end{equation}
instead of $m_A^2 \cos^2 2\beta + \epsilon / \tan^2 \beta$ and
$M_Z^2 \cos^2 2\beta + \epsilon$, where $m_A$ is the mass of the pseudoscalar
boson.  Assuming $m_t$ = 150 GeV and $\tilde m$ = 1 TeV, this means that
\begin{equation}
m_h < 120~ {\rm GeV}
\end{equation}
in this model, whereas $m_h < 115$ GeV in the MSSM.  There
is also the sum rule
\begin{equation}
m_{H^\pm}^2 = m_A^2 + {1 \over 2} M_W^2 \left( 1-{{2x} \over {1-2x}} \right)
\end{equation}
instead of the corresponding
$m_{H^\pm}^2 = m_A^2 + M_W^2$ in the MSSM.  In the limit of large $m_A$,
both models reduce to the standard model with $h$ as its one Higgs boson
while keeping their respective mass upper limits.
\vglue 0.6cm
{\bf\noindent 6. Outlook}
\vglue 0.4cm
If supersymmetry exists and future experiments discover two and only two
Higgs doublets at the electroweak energy scale, it does not mean
necessarily that the MSSM will be confirmed.  If this model with $f \neq 0$
or some other is found, then it will point to a larger theory such as
$\rm SU(2)_L \times SU(2)_R \times U(1)$ or some other at a higher energy
scale.
\vglue 0.4cm
{\bf\noindent 7. Acknowledgement}
\vglue 0.4cm
This work was supported in part by the U. S. Department of Energy under
Contract No. DE-AT03-87ER40327.
\vglue 0.6cm
{\bf\noindent 8. References \hfil}
\vglue 0.4cm

\end{document}